# RADIATION HAZARD OF RELATIVISTIC INTERSTELLAR FLIGHT

Oleg G. Semyonov[*]

*State University of New York at Stony Brook, Stony Brook, New York 11794, USA; osemyonov@ece.sunysb.edu*

**Abstract**

From the point of view of radiation safety, interstellar space is not an empty void. Interstellar gas and cosmic rays, which consist of hydrogen and helium nucleons, present a severe radiation hazard to crew and electronics aboard a relativistic interstellar ship. Of the two, the oncoming relativistic flow of interstellar gas produces the most intense radiation. A protection shield will be needed to block relativistic interstellar gas that can also absorb most of the cosmic rays which, as a result of relativistic aberration, form into a beamed flow propagating toward the front of the spaceship.

## 1. Introduction

"Interstellar travel may still be in its infancy, but adulthood is fast approaching, and our descendants will someday see childhood's end".[1] Ever since the revelation that all stars we see shining on the night's sky are ordinary dwellers of the Milky Way, just like our Sun, man has dreamed of journeys to other stellar systems. Most discussions on the possibility of interstellar flight have focused on analyzing propulsion systems capable of accelerating spaceships to a speed sufficient enough to reach nearby stars in a reasonable period of time.[1 - 3] Provided that the problem of propulsion is resolved, the decision making will be, apparently, based on a tradeoff between the cost of a mission and its practical expediency. The money involved seems to be enormous: for example, a flyby mission with a moderate speed of 0.3c, where c is the speed of light in vacuum, to one of the nearest stars could cost trillions or even tens of trillions of dollars,[4] and would require 50 - 100 years to accomplish. Relatively low-cost propulsion systems, like nuclear thrusters, are not feasible to accelerate a ship above 0.1c, which means that the missions will be unacceptably long. It is doubtful whether some commercial, national, or international entity would be willing to invest trillions in a mission that could take thousands of years to accomplish. Therefore, while the flights that utilize high relativistic speeds above 0.3c will be, apparently, even more expensive, they seem to be the only foreseeable option as no one will be, apparently, interested in financing missions that will take longer than man's lifespan. The main problem in achieving a relativistic velocity will, for now, remain that of propulsion, and it is far too early to suggest anything worthwhile for engineering design. Antimatter propulsion could one day provide the answer. However, it is beyond what we can create with existing technology and a thruster system utilizing it will be considerably costly, as well.[5] Nevertheless, a discussion of various approaches to interstellar travel and the analysis of physical problems arising on the way are still necessary, as it will prepare us for the future when the technology finally catches up with our dreams.

---

[*] E-mail address: osemyonov@ece.sunysb.edu

Propulsion is not the only issue we will face in trying to make interstellar travel possible. As a rule it is assumed that interstellar space is empty or, perhaps, contains a dilute and innocuous gas that might be useable for fuel replenishment or to produce drag for braking. In reality, this view of interstellar space is completely wrong. When a ship accelerates to a relativistic velocity above 0.3c, interstellar gas becomes a flow of relativistic nucleons, which, in itself, is nothing less than hard radiation bombarding the starship, its travelers, and all of the electronic equipment aboard. In addition, interstellar space contains high-energy cosmic rays and dust, all of which can present a huge problem if no proper protection is implemented. Thus, in order to consider the feasibility of manned or unmanned interstellar flight, we need first to evaluate radiation conditions aboard a ship. For comparison, a radiation dose obtained in a non-relativistic space module moving in interstellar space would be, approximately, 70 rems/year [6] whilst the safety level for a person is between 5 - 10 rems/year. The dose will, most likely, increase when accelerating to relativistic velocities. Irrespective of propulsion systems, future interstellar missions will not, apparently, be longer than ten to thirty years by a terrestrial clock, which means that starships will tend to utilize extreme relativistic velocities that would present significant radiation hazards onboard.

**2. Radiation hazard**

*2.1. Interstellar gas*

If the dimensionless velocity of a spaceship in a local system of coordinates, where interstellar gas can be considered at rest (thermal velocities of atoms are much less then a starship's speed), is $\beta = \mathcal{V}/c$, where $\mathcal{V}$ is the spaceship's velocity and c is the speed of light, relative energies of nucleons in an oncoming flow with respect to a coordinate system related to this ship will be: [7]

$$E = m_0 c^2 (\gamma - 1), \tag{1}$$

where $m_0$ is the mass of rest of nucleons and $\gamma = (1 - \beta^2)^{-1/2}$. Without shielding, a radiation dose produced by the oncoming relativistic flow of nucleons as a function of β can be estimated as:

$$D(rem/s) = 1.67 \cdot 10^{-8} \, Q \times n \times S \times \beta \times c \times H(\beta) \times d(\beta)/M, \qquad (2)$$

where Q is the radiation quality factor (Q = 10 for protons according to NRC Regulations Title 10, Code of Federal Regulations, § 20.1004 Units of radiation dose (2006)), $n(cm^{-3})$ is the concentration of interstellar gas, $S(cm^2)$ is the cross-section of human body (S ≈ $10^4$ $cm^2$, i.e. 1 $m^2$, is taken here for estimations), $H(MeV \, cm^2/g)$ is the stopping power of particles in human tissue, $d(g/cm^2)$ is the penetration depth of particles in tissue or the thickness of human body in the direction of particles' travel (whatever is less), and M(g) is the mass of an individual. The data for H and d as functions of energies of nucleons are taken from the NIST (National Institute of Standards and Technology) online database.

Interstellar gas consists of hydrogen (89%) with admixture of helium (10%) and a trace amount of heavier elements (1%) like carbon, oxygen, silicon, iron, etc. accreted in dust particles.[8, 9] Concentration of interstellar gas can vary between $10^4$ $cm^{-3}$ ($10^{10}$ $m^{-3}$) in the galactic clouds to less then 1 $cm^{-3}$ ($10^6$ $m^{-3}$) in the low-dense regions between the clouds. Fortunately, our Sun is located in an irregular cavity of low-density gas consisting of neutral hydrogen with concentration n(H) ~ 0.2 $cm^{-3}$ ($2 \times 10^5$ $m^{-3}$) and an ionized component having concentration n($H^+$) ~ 0.1 $cm^{-3}$ ($10^5$ $m^{-3}$); the nearest dense 'wall' is located at the distance of 170 light years in the direction of the center of our Galaxy.[8] Concentration of dust particles with their average internal density 2.5 $g/cm^3$ (2,500 $kg/m^3$) is ranged from $10^{-8}$ $m^{-3}$ in the low density regions to $10^{-5}$ $m^{-3}$ in the dense clouds and their masses vary from $10^{-17}$ kg to $10^{-20}$ kg.[9] The energies of nucleons (H, He) in the oncoming flow of interstellar gas versus a ship's velocity β and the corresponding doses of unshielded radiation are plotted in Fig.1a and b. The doses of unshielded radiation are extremely high: if a spaceship's speed exceeds 0.1c, an astronaut cannot be outside the hull without proper windward shielding even wearing a space suit. A break in the curve at β = 0.6c in Fig. 1b corresponds to a ship's velocity when the penetration depth of unshielded nucleons in tissue becomes equal to an average thickness of human torso d ≈ 35 cm (0.35 m).

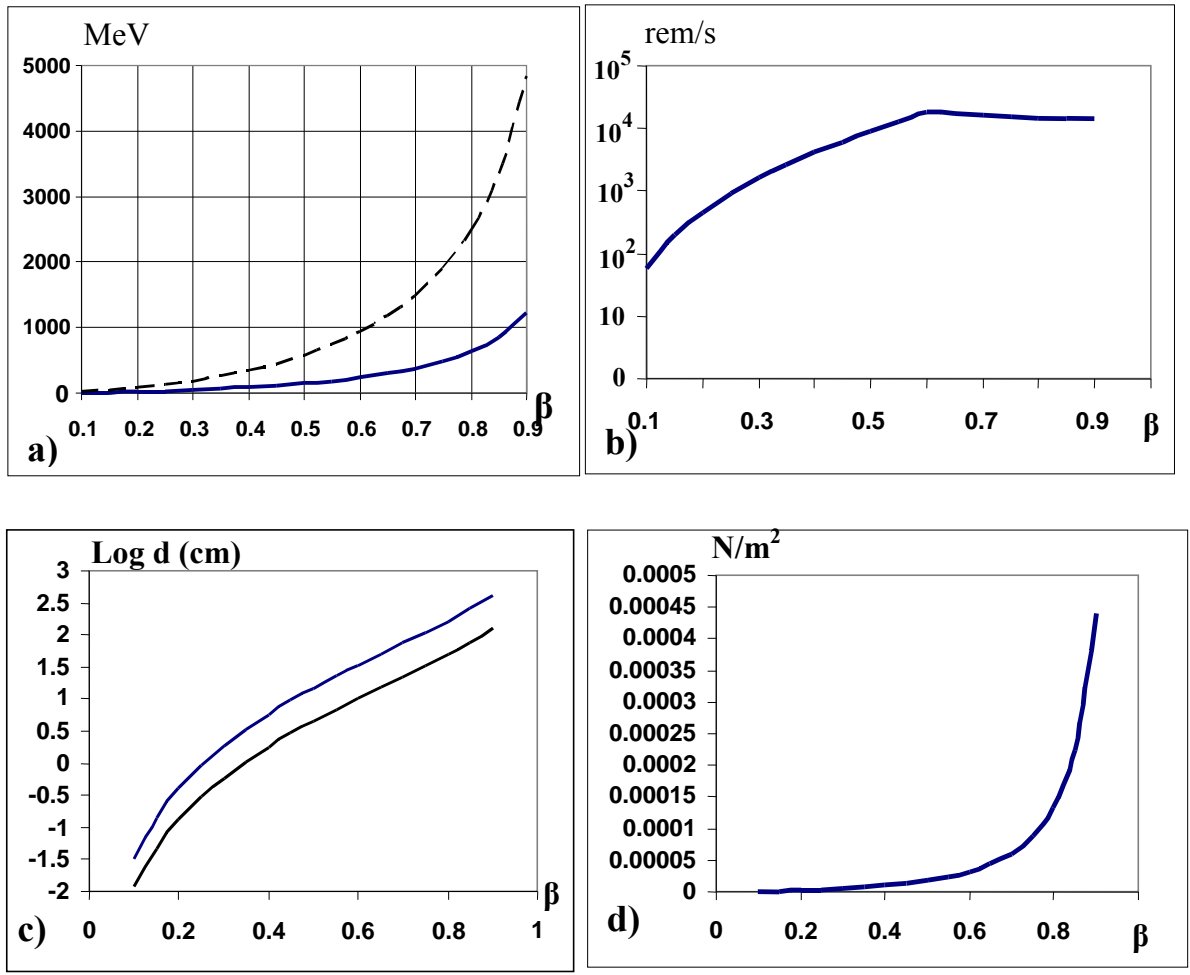

Fig.1. a) Energies of incident hydrogen (solid curve) and helium nuclei (dashed curve) of interstellar gas as functions of spaceship's velocity $\beta$; b) unshielded radiation dose (rems per second) versus $\beta$; c) penetration depths of protons in titanium (lower curve) and in tissue (upper curve); d) drag force imposed by the oncoming flow of interstellar gas.

No magnetic or electrostatic shielding is, apparently, applicable for protection against nucleonic radiation of the oncoming flow because of the presence of the neutral component in interstellar gas. Among the possible options of shielding, only two seem to be viable: a radiation-absorbing windscreen installed in front of a spaceship (or a ship's bow made of a radiation-absorbing material) or a combination of material and magnetic shielding (see details in Section 3.2). Penetration depth of nucleons in materials for a given cruising speed of a starship is quite distinct, so a shield, say, of double the penetration depth of hydrogen nuclei in the shield's material would provide reliable protection from direct radiation. Penetration depths of helium and hydrogen nucleons at a particular spaceship's speed are comparable in tissue and in other materials, for example, in titanium or aluminum (possible materials of choice for starships' hulls) despite some difference of their energies in the oncoming flow. For velocities up to 0.3c, a titanium 'windscreen' of 1 – 2 cm can provide sufficient protection, however it becomes dramatically thicker with acceleration above 0.3c (Fig. 1c) and reaches several meters at $\beta$ = 0.9. Shielding by water [5] could also be an option since astronauts would need a water supply onboard anyway. Placing a water tank (or an ice bulge) in front of a ship is advantageous in comparison with a shield made of metal or another solid material because it eliminates the damaging embrittlement of solids under intense nucleonic radiation: for a given cruising speed, the penetration depth of monoenergetic nucleons will be the same and a layer located near the penetration depth inside a solid shield will be largely damaged because all the nucleons deposit the bulk of their kinetic energy at the end of their penetration depth dislocating atoms from the lattice, weakening the material, and causing peeling or flaking. Below $\beta \sim 0.5$, a water tank of several tens of centimeters in thickness would be sufficient to reduce radiation down to a safe level, however cruising speeds closer to the speed of light would require tens of meters of water shielding, i.e. many tons of additional load.

As for the braking force $F(N/m^2) = d(mv)/dt = m_p n \beta^2 c^2 \gamma^3$ imposed by interstellar gas within the local low-density galactic cavity, it is fairy small (Fig. 1d); nevertheless, the drag can be quite significant in clouds, where concentration of gas is $10^4$ to $10^5$ times higher. The smaller clouds of elevated gas density within the local cavity [8] also can cause some transient elevations of radiation dose and drag force. Regarding the cosmic dust, two mechanical effects are anticipated: a) mechanical drag, which is approximately one order of magnitude smaller then the drag imposed by interstellar gas and b) deterioration of a frontal part of a hull or a frontal shield caused by intense micrometeorite bombardment with a rate exceeding 100 relativistic dust particles per square meter per second for $\beta > 0.3$ (their kinetic energy will exceed 0.1 J/particle), which would produce microexplosions inside the shield at the end of their passage through the shield's material, leaving, in addition, microholes to effectively increase its porosity.

*2.2. Cosmic rays*

Our galaxy is filled with high-energy nucleons that are generally thought to have originated in novae and supernovae.[19] Cosmic rays consist mostly of hydrogen and helium nuclei accelerated to relativistic energies and isotropically distributed over space by the galactic magnetic field.[10] Their energy maximum lies between 300 Mev and 1 GeV.

Consider a coordinate system S´ co-moving with a spaceship along the z-axis in a positive direction relative to a reference system S at rest. In analogy with the formula [11, 12] for the transformation of angles of incidence of photons observed from both frames, the following transformation rule between two coordinate systems S and S´ for the angles of incidence of relativistic particles $\theta$ and $\theta´$ with respect to the z-axis in spherical coordinates $\beta_p$, $\varphi$, $\theta$ can be found from Lorentz formula [7] for a velocity component collinear with a ship's velocity $\beta$ in the restframe S:

$$\beta'_p \cos\theta' = \frac{\beta_p \cos\theta - \beta}{1 - \beta_p \beta \cos\theta} \tag{3}$$

where $\beta'_p$ and $\beta_p$ are the dimensionless velocities of particles in the systems S´ and S, correspondingly, and $\theta'$ and $\theta$ are the angles of inclination of velocities of corresponding particles to the z-axis. The formula describes an effect of relativistic aberration of cosmic rays: the solid angles behind the ship are spreading out in the ship's frame S´ while the solid angles in front of the ship are squashing together as shown in Fig. 2.

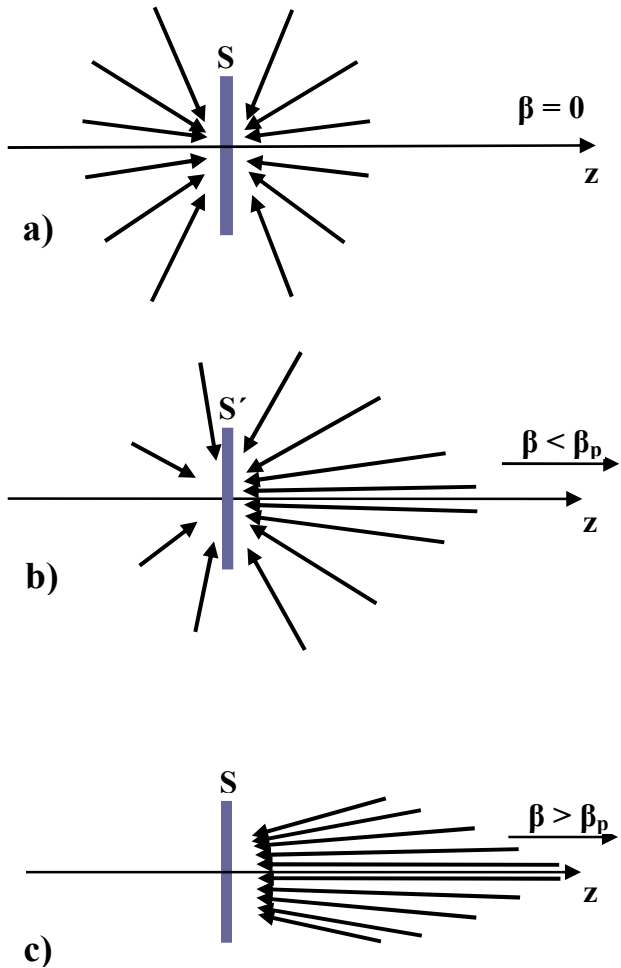

Fig. 2 Illustrative diagrams of angular distributions of originally monoenergetic and isotropic cosmic rays when a starship is: a) at rest with respect to a local reference system of coordinates S, b) moving with a relativistic speed $\beta$ below the average velocity of nucleons $\beta_p$ of cosmic rays in the reference frame S, and c) moving with a relativistic speed $\beta$ exceeding the velocity of nucleons $\beta_p$ in the reference frame S.

The flow of cosmic rays from the front hemisphere increases with β, while the flow from the rear hemisphere diminishes when the starship accelerates. Cosmic rays irradiating an area A in the restframe S are distributed isotropically as shown in Fig. 2a (monoenergetic particles in the restframe S are taken for illustration). The same particles observed from a spaceship moving along the z-axis in a positive direction with a relativistic velocity β will appear propagating at other angles as shown in Fig. 2b and c. When a spaceship is at rest in a local restframe S related to stars or gaseous objects in a particular part of the Galaxy, both surfaces of a ship's cross-section are equally irradiated by cosmic rays (Fig. 2a). With β increasing, more and more cosmic rays of correspondingly increasing energy irradiate the frontal surface of area A´ = A in the frame S´ (to an extent, this phenomenon is analogous to an oncoming flow of snowflakes observed from a moving car) while the flux of cosmic rays from the rear decreases together with their kinetic energies (Fig. 2b). At $\beta \geq \beta_p$, the flow incoming from the rear hemisphere disappears completely and all cosmic rays come from the front hemisphere (Fig. 2c). In the restframe S, concentration $n$ of cosmic rays in the vicinity of the Sun was measured ~ $10^{-3}$ m$^{-3}$ and their distribution in energy can be expressed as $\mathcal{N}$(m$^{-2}$ s$^{-1}$ sr$^{-1}$ GeV$^{-1}$) = $1.8 \times 10^4$ $E_p^{-2.7}$ for $E_p$ > 1 GeV.[10] Approximating it by a flow of monoenergetic nucleons with an average energy of 1 GeV, a total flow of nucleons is, approximately, 1.8 nucleons per square centimeter per steradian per second, which yields a total flow N ≈ 11 nucleons per square centimeter per second (110,000 per square meter per second) from all directions through the ship's cross-section perpendicular to the direction of cruising. For a non-relativistic flight, the unshielded absorption dose can be estimated as:

$$D = 0.53 \times N \times H(\text{MeV/cm}) \times d(\text{cm}) \times S(\text{cm}^2)/M(\text{g}) \approx 35 \text{ rads/year}, \qquad (4)$$

where H ≈ 2 MeV/cm (200 MeV/m) is the stopping power of 1GeV cosmic rays in tissue (water), d ≈ 30 cm (0.3 m) is the average thickness of human torso, S ≈ $7 \times 10^3$ cm$^2$ (0.7 m$^2$) is the average cross-section of human body, M = 70000 g (70 kg) is the approximate mass of an adult astronaut. Thus, the equivalent dose of unshielded radiation Q×D can reach 350 rems/year, which is quite discouraging because a dose of 100 rems accumulated during a relatively short interval can cause acute radiation sickness and possible death from cancer with a probability of 10%.

Using, in addition to (2), the Lorentz formula for a transverse component of velocity of the nucleons in cosmic rays, the functions $\beta'_p(\theta)$ and $\cos\theta'(\theta)$ can be found separately:

$$\beta'_p(\theta) = \left[ \frac{(1-\beta\beta_p Cos\theta)^2 - (1-\beta^2)(1-\beta_p^2)}{(1-\beta_p\beta Cos\theta)^2} \right]^{1/2} ; \qquad (5)$$

$$Cos\theta' = \frac{\beta_p Cos\theta - \beta}{[(1-\beta_p\beta Cos\theta)^2 - (1-\beta^2)(1-\beta_p^2)]^{1/2}} . \qquad (6)$$

These are the general expressions for any type of particles; in the case of photons ($\beta_p = 1$), they are reduced to the well-known formulae for relativistic transformation of angles of propagation of light rays (relativistic aberration of light).[11, 12] The angles of incidence $\theta'$ observed from a spaceship as functions of their angles of incidence $\theta$ in the restframe S and the velocities $\beta'_p$ in the frame S´ for various $\beta$ are shown in Fig. 3. For greater clarity and simplicity, cosmic rays are approximated by an isotropic and monochromatic (1 GeV) flow of protons in the restframe S.

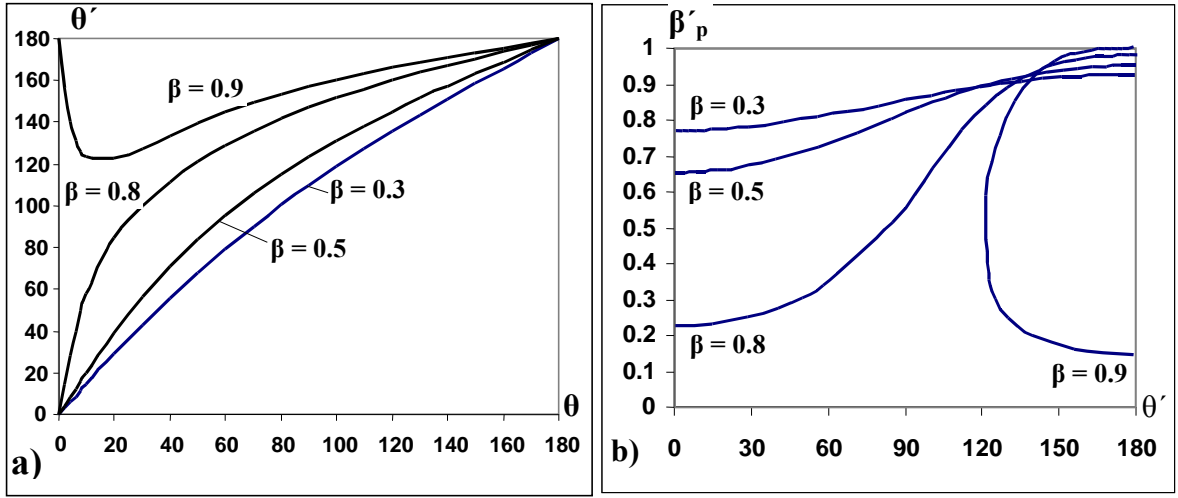

Fig. 3 a) Angles (in degrees) of incidence of cosmic rays $\theta'$ observed from a relativistic spaceship as functions of their angles of incidence $\theta$ in the restframe S. b) Velocities $\beta'_p$ of nucleons observed from a starship as functions of angles $\theta'$ for different cruising velocities $\beta$, assuming isotropic distribution of monoenergetic cosmic rays with $\beta_p = 0.87$ ($E_p = 1$ GeV) in the restframe S. When $\beta > \beta_p$ (the upper curve in Fig. 3a and the right curve in Fig. 3b), no cosmic rays are incident from the rear; even those cosmic rays, which were incident at small angles $\theta$ and moved in pursuit of the ship, are now moving toward the ship with relatively low velocity $\beta'_p$.

In analogy with the deductions of McKinley [12] and Weiskopf at al. [13] for photons, and assuming equality of total counts (total number of nucleons per unit of ship's cross-section perpendicular to the ship's velocity) measured in both coordinate systems for the correlated time intervals, a total flow of monoenergetic (E = 1 GeV) cosmic rays on a unit area perpendicular to the direction of movement (z-axis) as a function of $\theta'$ in the spaceship frame S´ can be expressed as:

$$\mathcal{N}'(\theta') = \mathcal{N}\gamma \frac{\beta'_p}{\beta_p} \frac{d\Omega}{d\Omega'}(1 - \beta\beta_p Cos\theta) = 1.8\,\gamma \frac{\beta'_p}{\beta_p}\left[\frac{d(Cos\theta')}{d(Cos\theta)}\right]^{-1}(1 - \beta\beta_p Cos\theta), \quad (7)$$

where $\gamma$ is the time dilation factor, the ratio $\beta'_p/\beta_p$ expresses a change of flow in a given direction observed from a spaceship due to alteration of relative velocities of the corresponding nucleons, and the expressions in brackets describe an additional change of flow due to narrowing of the solid angles in front of the ship and their spreading out behind the ship, where:

$$\left[\frac{d(Cos\theta')}{d(Cos\theta)}\right] = \frac{(1-\beta^2)\beta_p^2(\beta_p - \beta Cos\theta)}{[(1-\beta_p\beta Cos\theta)^2 - (1-\beta^2)(1-\beta_p^2)]^{3/2}} \quad (8)$$

A total flow of cosmic rays from both hemispheres per unit area of a spaceship's cross-section perpendicular to the direction of movement per unit time according to a spaceship's clock is:

$$N' = \iint \mathcal{N}'(\theta')\,Sin\theta'Cos\theta'd\varphi'd\theta' = -2\pi \int_{-1}^{-1} \mathcal{N}'(\theta')\,Cos\theta'd(Cos\theta') =$$
$$= -3.6\pi\gamma \int_{1}^{-1} \frac{1}{\beta_p}(\beta_p\,Cos\theta - \beta)\,d(Cos\theta). \quad (9)$$

(To avoid an error resulting from oddness of the cosine function on an interval 0 - 180°, this integral should be taken separately over the front hemisphere and the rear hemisphere in the frame S´ with subsequent summation of the results; it corresponds to integration over $\theta$ (last integral) first between 1 and $\beta/\beta_p$ and then between $\beta/\beta_p$ and -1 with subsequent summation of the absolute values).

A total equivalent dose of unshielded cosmic rays in the spaceship's frame S´ can be calculated from:

$$D(rems/year) = 2\pi \times 5.3 M^{-1} \int \mathcal{N}'(\theta')\,H(\theta')\,d(\theta')\,S(\theta')\,d\theta'. \quad (10)$$

Taking into account a dramatic reduction of low-$\beta'_p$ (therefore, low-energy) nucleons from the rear hemisphere (Fig. 4 b), when $\beta$ increases, and an approximate constancy of stopping power H of protons when their energy exceeds 0.5 GeV (H ≈ 2 MeV/cm in tissue), the unshielded dose can be estimated as D = 5.3×N´×H×d×S/M ≈ 32×N´ rems a year, where N´(cm$^{-2}$ s$^{-1}$ sr$^{-1}$ GeV$^{-1}$) as a function of $\beta$ is plotted in Fig. 4a.

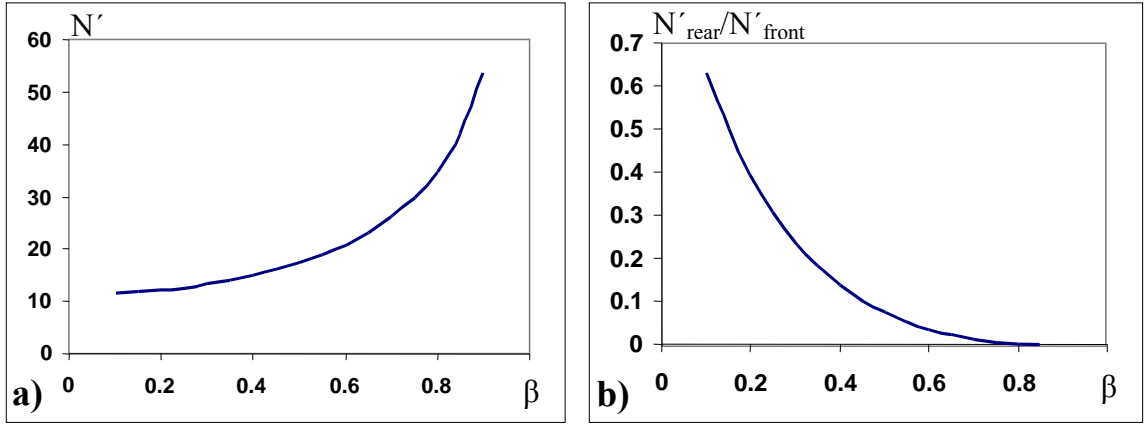

Fig. 4. a) Total flow of monoenergetic (1 GeV) cosmic rays per second through one square centimeter of a cross-section perpendicular to the direction of movement; b) ratio of fluxes of cosmic rays from the rear hemisphere and from the front hemisphere as a function of $\beta$.

The flow of cosmic rays from the rear drops down to zero when the ship's speed increases to $\beta = \beta_p$ (Fig. 4b); besides, their kinetic energy decreases dramatically when $\gamma'_p = 1/(1 - \beta'^2_p)^{1/2} \to 1$. Correspondingly, the flow of cosmic rays bombarding the frontal surface increases, the nucleons gain their kinetic energy, and the oncoming flow becomes increasingly narrower (relativistic beaming). This frontal beaming of cosmic radiation has a positive effect: the higher relativistic velocities of starships become advantageous because a frontal shield, which is needed anyway for protection against hard radiation of the relativistic flow of interstellar gas, will protect it against a major portion of cosmic rays, too. Obviously, cosmic rays of originally lower energies are subjected to stronger relativistic beaming and their capture by the frontal shield will be even more effective.

**3. Discussion**

3.1 *Material shielding.* Three nucleonic components in interstellar space are hazardous for starships cruising with relativistic velocities: a neutral component of interstellar gas, an ionized component, and cosmic rays; in addition, interstellar dust can also produce radiation hazard and mechanical damage. Interstellar gas is virtually innocuous for non-relativistic spaceships, but it will be one of the major concerns when a spaceship accelerates to a relativistic speed because it acts as hard radiation produced by the high-energy nucleons. This nucleonic radiation is capable of piercing through the hull and penetrating the crew quarters. Because of relatively higher density in comparison with cosmic rays, the dose of unshielded radiation caused by interstellar gas is extremely high (thousands or hundreds of thousands rems per second, which is comparable with the conditions in the cores of nuclear reactors), and an astronaut exposed to this radiation without a proper windward shielding will be virtually instantly killed by radiation. Shielding seems to be easy for ship's velocities below 0.3c; a titanium or aluminum hull of 1 to 2 cm in thickness can provide sufficient protection against the oncoming flow of nucleonic radiation. However, when the ship's speed is close to the relativistic barrier c, say 0.8 – 0.9c, a windward shield of several meters of titanium or, more practically, tens meters of water will be needed. The showers of secondary radiation (muon and gamma) produced by the high-energy nucleons in such a shield will also contribute to the radiation dose inside the ship. This is because of their high penetration ability, and an additional layer of high-Z material must be mounted to absorb this secondary radiation.

Cosmic rays consist of high-energy protons and alpha-particles, which means that the radiation hazard is tangible both for non-relativistic and relativistic space flights. Strictly speaking, a complete shielding against cosmic rays would require something analogous to Earth's atmosphere which can be implemented by, for example, a round shell of water of 5 meters in thickness.[6] This will be, perhaps, not a desirable solution both for interplanetary and interstellar flights due to the inevitable increase in mass. Even a water shell of 1 m in thickness, satisfying the radiation safety standard, could be excessively heavy. Besides, an additional layer of dense material will be needed to absorb the highly penetrating secondary gamma and muonic radiation, which inevitably enlarges the mass of the ship. If NASA's limit of 400 rads per individual during his duty, meaning the doubled probability to develop cancer, will be accepted for starflights, the thickness of a material shield and, therefore, its mass can be significantly reduced for the short-term missions (1 to 5 years); notwithstanding, the long-term interstellar travels will require more effective shielding. Advantageously, cosmic rays become increasingly beamed for higher relativistic velocities of spaceships. A frontal shield, which is needed anyway to protect the crew from the relativistic nucleonic 'wind' of interstellar gas, will also absorb a lion's share of cosmic rays. This is providing that the shield is designed to effectively attenuate the flow of cosmic rays together with the oncoming flow of interstellar gas, despite the increased total number of cosmic rays of increased energy (Fig. 4a) per unit area per unit time.

Even the relatively moderate relativistic velocities $\beta \leq 0.9$, considered here, impose significantly higher requirements on the material shielding of interstellar ships in comparison with interplanetary modules. The situation worsens dramatically for ultra-relativistic velocities. For example for $\beta = 0.995$, all cosmic rays will come from the front and their kinetic energy will be near 10 GeV; the penetration depth of protons of this energy will be ~ 40 m in water and ~ 10 m in titanium. Obviously, such a frontal shield falls out of reasonable scale.

3.2 *Combination of material and magnetic shielding.* The strength of a magnetic field needed for the complete reflection of electrically charged high-energy particles with energies ranging from several hundreds MeV to several GeV is ~ 20 Teslas, [6] and such a strong magnetic field is hazardous for both the crew and electronics. In principal, a suitable ferromagnetic shell covering the crew quarters can protect against magnetic field, if an inevitable increase of mass can be tolerable. A better solution is to put the living quarters in a region of zero magnetic field, [6] for example, if a ship is built in the form of a hollow torus with its external wall covered by superconducting current-carrying coils mounted along the generatrix. As it has been mentioned, magnetic shielding alone is not a solution for protection against the relativistic flow of interstellar gas because of presence of a neutral component. An effective solution could be found in combination of both material and magnetic shielding. With a combined shield, the thickness and mass of a frontal shield would be decreased dramatically both for moderate and ultra-relativistic velocities while the potentially harmful action of magnetic field on crew and electronics could be eliminated completely without significant infringement on architecture and mass of a starship. It could, for example, consist of two parts: a relatively thin solid disk (umbrella) at some distance in front of a starship, and superconductive magnetic coils mounted behind the disk. The coils generate an azimuthal magnetic field with magnetic lines perpendicular to the ship's velocity. The disk can be virtually transparent to the oncoming nucleons and atoms; its task is to strip the neutral atoms from their electrons and produce a flux of completely charged particles behind. Electron stripping of high-energy relativistic atoms by means of metallic foils or gas/plasma sheets is a well-known and commonly used method in ion accelerators.[14] After the electron stripper, all the charged particles of the oncoming flow would be deflected away from the starship by a magnetic field of proper strength generated by the superconductive coils. Such a combined shield is more attractive because the electron stripper can be of relatively low mass and the magnetic system can also be lightened considering that a relatively low magnetic field (0.1 – 1 T for the coils of 1 to 10 m in size) would be needed for the deflection of the high-energy nucleons away from the ship. However, such the system has one drawback: a negative charge will be accumulated on the stripper and, because of that, the starship will become increasingly negatively charged due to an accumulation of electrons and following deflection of the stripped positive ions to the surrounding space. To avoid this charge accumulation, which leads to appearance of a megavolt electric field around the ship causing the charged particles of surrounding gas to accelerate toward the ship hull and to create an additional severe radiation hazard, the area inside the magnetic coils with a magnetic field can be filled with a lightweight liquid dielectric, say, cryogenic hydrogen or helium, as it will be needed for the superconductive magnets anyway. Thus, a transverse magnetic field of sufficient strength in a tank of cryogenic liquid will force all the charged particles of the oncoming flow to gyrate inside the tank. The shield's thickness h can be significantly smaller in comparison with the penetration depth of nucleons in this dielectric material:

$$h \sim 2R = \frac{2m_0 \beta'_p C}{qB\sqrt{1-\beta'^2_p}} \quad (11)$$

where R is the radius of gyration of charged particles for a given velocity $\beta'_p$, $q$ is the charge of particles, and B is the magnetic field. For example, a tank needed for protection against the high energy H and He-atoms of the oncoming interstellar gas flow will be ≈ 1 m in thickness, if B = 10 T and $\beta'_p$ = 0.8 (starship speed $\mathcal{V}$ = 0.8c), which is significantly smaller than the penetration depth d ~ 10 m of the nucleons in liquid hydrogen. Such a system will also be able to effectively attenuate beamed cosmic rays of moderate energy and to deflect their high energy portion away from the ship. The problem of charge accumulation is significantly simplified because virtually all the positively charged nucleons of oncoming gas, including the stripped ones, would be absorbed in the tank to compensate the accumulation of negatively charged electrons in the electron stripper. In addition, secondary (muon and gamma) radiation would be distributed over $2\pi$ angles reducing its portion directed to the ship thus facilitating protection against it. A toroidal cryogenic tank with its enveloping magnetic coils mounted in front of a spaceship, which, in its turn, has the form of a hollow cylinder, looks preferable, because all of the magnetic field will be completely enclosed inside the toroidal tank with the absorbing liquid. Such a configuration will also allow us to implement rotation of the ship around its axis of symmetry to increase its spatial stability and to create artificial gravitation. Embrittlement of the tank's frontal wall or the stripper by the oncoming nucleons of interstellar gas and their sputtering by dust are among the problems that remain to be solved. Obviously, a shielding system designed for protection within the local low-density cavity in our Galaxy could be insufficient in the high density clouds. Fortunately, our Sun is located in a low density cavity of several hundred light years in size, and several thousand neighboring stars are not blocked by dense clouds. Still, interstellar navigation charts and maps of interstellar clouds will be needed for laying out a safe course both in the cavity and in the low-density tunnels [8] beyond it.

3.3 *Radiation hazard for electronic components*. Every high-energy nucleon passing through an electronic component inevitably produces free electrons, i.e. deposits some electric charge in it, often resulting in parasitic signals and causing bits to flip, latch up, or burn out in computer memory. The deposition of charge can "upset" the memory circuits. The so-called Single Event Effects (SEEs) and Single Event Upsets (SEU) in computer memory were intensively studied in high-altitude avionics, [15] on satellites, and in multiple laboratory tests.[16, 17, 18] The upset rate of a particular part of electronic equipment caused by cosmic radiation in the vicinity of our planet can vary from ten per day for commercial one-megabit RAMs to 1 every 2800 years for radiation-hardened one-megabit RAMs (radiation-hardened component is a device specially designed to resist nucleonic radiation). Current estimates of SEEs in aviation at 10-20,000 meters altitude are around $10^{-9}$ to $10^{-8}$ per bit per hour.[15] There are two other effects that can cause degradation of electronics. Total Dose Effect is the change of electrical properties of components upon their exposure to radiation. Displacement Damage occurs when nucleons slow down and nearly come to rest at the end of their penetration depth. As a result, they knock silicon atoms out of their proper locations in crystal lattice, creating defects in crystal structure capable of trapping conduction electrons.

The measured SEE in CMOS (complementary metal-oxide semiconductors – a major class of integrated circuits) [16] is approximately 2.6 $10^{-7}$ cm$^2$/bit (2.6 $10^{-11}$ m$^2$/bit) for particles' LETs (Linear Energy Transfer) above 20 MeV·cm$^2$/mg (2,000 MeV·m$^2$/kg) with a dramatic decrease below 14 MeV cm$^2$/mg (1,400 MeV·m$^2$/kg). The laboratory tests of various electronic components irradiated by protons and heavy ions were performed by LaBel at al.[17, 18] SEEs, SEUs, and SELs were detected virtually in all devices bombarded by heavy ions. Some showed SEEs and SEUs under proton irradiation with a characteristic cross-section of $10^{-13}$ cm$^2$/bit ($10^{-17}$ m$^2$/bit) or lower while the cumulative effects such as degradation of current transfer ratio, reference voltage degradation, functional failure, and displacement damage were commonly observed under protonic fluences above $10^{11}$ cm$^{-2}$ ($10^{15}$ m$^{-2}$) . Therefore, all electronic devices aboard relativistic interstellar ships must be properly shielded from the oncoming 'wind' of high-energy nucleons of interstellar gas; the flow of hydrogen atoms in the local cavity will exceed $3\times10^9$ cm$^{-2}$ s$^{-1}$ ($3\cdot10^{13}$ m$^{-2}$ s$^{-1}$) for the ship's velocities above 0.3c, which means that every unshielded electronic component will degrade to an inoperable condition in minutes. Cosmic rays do not seem to be of great concern, especially for the radiation-hardened devices with malfunction rate of ~ $10^{-9}$ – $10^{-10}$ errors/bit-day. Despite increased flow and energy of particles of beamed cosmic rays at relativistic velocities of starships, these cosmic rays will be largely absorbed or deflected away from a ship by the combined windward shielding system already in place to deal with relativistic interstellar gas.

## 4. Conclusion

From the point of view of radiation safety, interstellar space is not an empty void. Three nucleonic components in space are hazardous for crew and electronics on relativistic spaceships: neutral and ionized components of interstellar gas, and cosmic rays. The most dangerous is interstellar gas, which acts as a flow nucleonic radiation bombarding a relativistic starship. Radiation flux is extremely high even for moderate relativistic velocities and, therefore, proper windward shielding is a necessity. The presence of a neutral component in interstellar gas excludes the implementation of magnetic shielding alone. Either a sufficiently thick material shield or a significantly lighter combination of an electron stripper followed by a magnetic system enveloping a cryogenic tank ought to be mounted in front of the starship.

Isotropic cosmic rays are subjected to relativistic beaming when a starship is moving with a relativistic speed. For the ship's velocities closer to the speed of light, most of cosmic rays form into a beam directed toward the front of the spaceship. While they do present a hazard, they can be easily absorbed or deflected by a frontal shielding system that is required anyway protecting the crew and electronics against the hard radiation of the oncoming flow of interstellar gas. Cosmic dust will also contribute to the radiation hazard, because the dust particles are actually lumps of high-energy nucleons at relativistic velocities. A serious problem will be the sputtering of a ship's bow or a radiation shield by the relativistic dust particles. Nevertheless, the shielding of relativistic starships from hard ionizing radiation produced by interstellar gas and cosmic rays does not seem to be far beyond existing technology.

**References**

[1] E. Mallove, and G. Matloff, The Starflight Handbook, John Wiley and Sons, New York, 1989, pp. 71-105.

[2] R. L. Forward, Roundtrip interstellar travel using laser-pushed lightsails, J. Spacecrafts and Rockets 21 (1984) 187-195,.

[3] R. L. Forward, Feasibility of interstellar travel, J. of the British Interplanetary Society 39 (1986) 377-400.

[4] C. Mileikowsky, Cost considerations regarding interstellar transport of scientific probes with coasting speeds of about 0.3c, 45th Congress of the International Astronautical Federation (1994), paper IAA-94-655.

[5] D. Leonard, Reaching for interstellar flight, Space News (2003)
http://www.msnbc.msn.com/id/3741674/.

[6] E. Parker, Shielding space explorers from cosmic rays, Space Weather 3 (2005) S08004.

[7] W. Benenson, J. W. Harris, H. Stocker, H. Lutz, (editors), Handbook of Physics, Springer-Verlag, New York, 2001, p. 143-147

[8] P.C. Frisch, The galactic environment of the Sun, Journal of Geophysical Research 105 (2000) 10279; see also http://spaceflightnow.com/news/n0305/30map3d/

[9] I. Mann, K. Hiroshi, Interstellar dust: properties derived from mass density, mass distribution and flux rates, Journal of Geophysical Research 105 (2000) 10317.

[10] J. A. Simpson, Elemental and isotopic composition of the galactic cosmic rays, Ann. Rev. Nucl. and Part. Sci. 33 (1983) 323-382.

[11] C. Møller, The Theory of Relativity, Clarendon Press, Oxford, 1972.

[12] J. M. McKinley, Relativistic transformation of solid angle, Am. J. Phys. 48 (1980) 612-614.

[13] D. Weiskopf, U. Kraus, and H. Ruder, Searchlight and Doppler effects in the visualization of special relativity: A corrected derivation of the transformation of radiance, ACM Transactions of Graphics, 18 (1999) 278-292.

[14] G. D. Alton, R.A. Sparrow, R. E. Olson, Plasma as a high-charge-state projectile stripping medium, Phys. Rev. A 45 (1992) 5957- 5963.

[15] E. Normand, Single event effects in avionics, IEEE Transactions on Nuclear Science 43 (1996) 461-474

[16 F. Faccio, K. Kloukinas, A. Marchioro, T. Calin, J. Cosculluela, M. Nicolaidis, R. Velazco, Single effect in static and dynamic registers in a 0.25 mkm CMOS technology, IEEE Transactions on Nuclear Science 46 (1999) 1434-1439.

[17] K. A. LaBel, P.W. Marshall, J.L. Barth, R.B. Katz, R.A. Reed, H.W. Leidecker, H.S. Kim, C.J. Marshall, Anatomy of anomaly: investigation of proton-induced SEE test results for stacked IBM DRAMs, IEEE Transactions on Nuclear Science 45 (1998) 2898-2903.

[18] K. A. LaBel, P.W. Marshall, C.J. Marshall, M. D'Ordine, M. Carts, G. Lum, H.S. Kim, C.M. Seidleck, T. Powell, R. Abbott, J. Barth, E.G. Stassinopoulos, Proton-induced transients in octocouplers: in-flight anomalies, ground irradiation tests, mitigation, and implications, IEEE Trans. Nuclear Science 44 (1997) 1885-1892; R. A. Reed, P.W. Marshall, A.H. Johnston, J.L. Barth, C.J. Marshall, K.A. LaBel, M. D'Ordine, H.S. Kim, M.A. Carts, IEEE Transactions on Nuclear Science 45 (1998) 2833-2841.

[19] D. Perkins, Particle Astrophysics, Oxford University Press, Oxford, 2003